# Analyzing the Impact of Strategic Bidding on the Reserve Capacity via a Bi-Level Model

Yun Xu, Yunxiao Bai, Yunyong Zhang, Jiqun Guo, Kaijun Xie
Guangdong Power Exchange Center Co., Ltd
Guangdong, China
1192678054@qq.com, 516486451@163.com,
121769438@qq.com, 17862996392@163.com,
xiekaijun@foxmail.com

Peng Wang, Xuelin Wang, Rusheng Zhao
Sichuan Energy Internet Research Institute
Tsinghua University
Sichuan, China
peng.wang@alumnos.upm.es, 1696955301@qq.com,
zrszrs321@gmail.com

*Abstract*—The growing integration of renewable energy sources necessitates adequate reserve capacity to maintain power balance. However, in market clearing, power companies with flexible resources may submit strategic bids to maximize profits, potentially compromising system reserves. This paper examines the effects of such strategic behavior by modeling the market as a bi-level problem. The upper level represents a strategic company aiming to maximize profit, while the lower level simulates the system operator clearing the market based on submitted offers. To enable duality-based solution methods, we approximate unit commitments with a continuous reserve capacity calculation. Case studies indicate that, in an imperfectly competitive market, more units are incentivized to operate, enhancing system reserves. However, some units go online mainly for profit, ultimately raising electricity costs for consumers. These findings highlight the importance of market design in managing the trade-off between reserve adequacy and economic efficiency in the presence of strategic bidding behavior.

*Index Terms*--Bi-level optimization, strategic bidding, market power, reserve capacity

## I. Introduction

In order to achieve carbon neutrality by 2060, a large number of Renewable Energy Sources (RES) are being connected to the grid in China [1], gradually replacing traditional flexible units, such as Synchronous Generators (SGs). This inevitably reduces the reserve capacity of systems, making it difficult to maintain the balance between supply and demand under the impact of RES such as wind and solar power generation [2]. In the energy market, in order to make a profit, power companies that manage certain SGs may report a higher marginal cost in the market clearing [3], which will directly affect the operation of the remaining units and indirectly exert impact on the system reserve.

To capture the strategic behaviors of power companies, i.e., misreport marginal prices, and then examine the influences they may have, i.e., market power [4], this work adopts a bi-level model to simulate an imperfectly competitive market, in which the energy biddings may be strategic. Bi-level model is a methodology in the game theory, where the strategic player at the Upper Level (UL) makes decisions, and the followers at the Lower Level (LL) accept them and respond. This method has been successfully applied in the similar studies [3,5], revealing how market power manifests.

The typical method to solve the bi-level model is to derive the equivalent Karush-Kuhn-Tucker (KKT) conditions of the LL problem and then solve it in conjunction with the UL problem, i.e., to model the entire problem as a single-level manner. However, this strictly requires that the entire problem is continuous and convex, otherwise the dual variables cannot be defined and the KKT conditions cannot be derived [3]. This means that problems involving Unit Commitment (UC) property cannot be solved directly, and new solution algorithms or model assumptions need to be proposed to solve problems in the UC context. Literature [6] proposes an iterative solution algorithm to address the UC issue, the core of which is the column-and-constraint generation. Another algorithm is to relax the binary variables, derive the dual problem of the LL problem, then restore the discreteness of the variables, and use the penalty function method to minimize the dual gap [3], and eventually form the final single-level problem with the UL problem. This method has also been applied widely. However, the above methods require complex mathematical derivations, and the final solution is still not global optimal, but a suboptimal solution based on an error given in advance. In order to make the bi-level model non-convex and continuous, and to define the spinning reserve based on the on/off state, we propose an assumption to calculate the reserve capacity offline. Although this removes the startup and shutdown cost and no-load cost, making the model conservative, the solution is globally optimal and the model is easy to construct and understand.

The remainder of the paper is organized as follows: Section II describes the offline calculation method of reserve

This work was supported by the Science and Technology Project of Guangdong Power Exchange Center Co., Ltd. No: GDKJXM20250087. (Corresponding address: peng.wang@alumnos.upm.es)

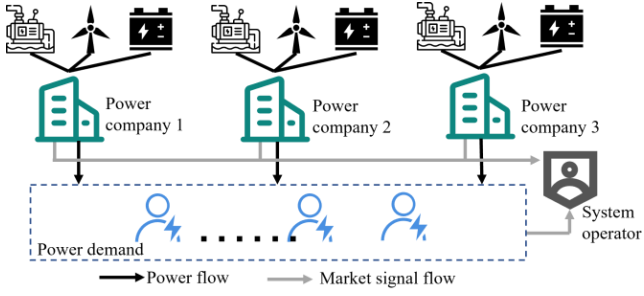

Figure 1. Structure of the energy market considered.

capacity and the test process. Section III builds the bi-level model and the quantification method of market power. Section IV carries out case studies. Section V summarizes this work.

## II. DESIGN OF TEST FRAMEWORK

In this section, we first explain how to calculate the reserve capacity in an offline manner based on the dispatch results. Then, logistics of simulation for comparing various scenarios to capture the market power are clarified.

### A. Spinning Reserve Capacity Computation Method

The reserve capacity is often required to be provided by the units that have been online for a fast response, i.e., spinning reserve, which means that there is a clear UC status, and binary variables are consequently introduced into the model. However, for a bi-level model, the introduction of discrete variables is undoubtedly asking for trouble, which will lead to the infeasibility of the KKT optimality conditions and model hard to be solvable [7].

Therefore, to develop a bi-level problem with spinning reserve modelling, we assume that the lower bound of the output of SGs can be 0 to indicate that they are offline, thus having no spinning reserve. Vice versa, when they are turned on and generating electricity, the remaining capacity can be served as the reserve. Here, we consider two different mathematical formulas to express two types of the reserve. The logic statement of this assumption is expressed as:

$$R_{t,\max} = \begin{cases} 0, & \text{if } P_{\text{SG},t} = 0 \\ P_{\text{SG,max}} - P_{\text{SG},t}, & \text{if } P_{\text{SG},t} \neq 0 \end{cases} \quad (1.1)$$

$$R_t = \begin{cases} 0, & \text{if } P_{\text{SG},t} = 0 \\ \min\left\{P_{\text{SG,max}} - P_{\text{SG},t},\ P_{\text{SG}}^{\text{up}}\right\}, & \text{if } P_{\text{SG},t} \neq 0 \end{cases} \quad (1.2)$$

where $P_{\text{SG,max}}$ is the maximum generation of SGs (MW). $P_{\text{SG}}^{\text{up}}$ is the ramp-up rate of SGs (MW/h). $P_{\text{SG},t}$ is the output of SGs at period $t$ (MW). (1.1) means the maximum possible reserve capacity, i.e., ignoring the technical limitations of ramping. (1.2) takes into account the actual ramping constraints to consider the generation that can be dispatched up within the hour as the available reserve capacity.

The maximum possible reserve (Type-I) reflects the static regulation capability of the unit, while the actual reserve capacity (Type-II) reflects the dynamic regulation capability of the unit under time constraints. The difference between the two lies in whether the ramp rate and response time are considered. The former requires no time limit or a sufficiently long time, indicating the potential of spinning reserve, i.e., the gap from current output to the rated capacity. The latter measures the ability of spinning reserve to respond quickly in a short period of time to participate in scheduling.

### B. Simulation Framework

Before describing the simulation process, the market structure in this paper is introduced, as shown in Fig. 1. Three power companies with some flexible resources: SGs, Wind Turbines (WTs), Battery Storages (BSs), participate in the electricity market clearing. The System Operator (SO) collects market information of supply and demand including hourly supply-demand gap, energy price offers from power companies, then clears the market.

In the market considered, we design two different clearing scenarios, i.e., the non-strategic case and strategic cases. In the former case, the clearing problem is a single-level problem, which means that each power company is not selfish and reports the true marginal generation cost to the SO. In the latter case, there is one strategic bidding player who may submit bids to the SO that are higher than the true prices in order to selfishly maximize their own interests. Based on the dispatch results in these two scenarios, we analyze the impact of strategic player on the market and system reserve capacity.

## III. STRATEGIC BI-LEVEL MODEL FORMULATION

In this section, the strategic power company maximizes its profits through strategic bidding in the UL problem, while the SO at the LL tries to minimize the operating cost of the entire system. Finally, how to quantify the market power of the strategic player is modeled according to the Lerner index.

### A. Upper Level Model

The profits of the strategic power company are made of two parts: the revenues of market clearing minus the operating costs. The mathematical model is expressed as:

$$\max_{k_t} \sum_t \left[ \lambda_t P_{\text{SG},t}^{\text{str}} - O_{\text{SG}}^{\text{str}} P_{\text{SG},t}^{\text{str}} - O_{\text{BS}}^{\text{str}} {P_{\text{BS},t}^{\text{str}}}^2 \right] \quad (2.1)$$

subject to:

$$1 \leq k_t \leq k_{\max} \quad (2.2)$$

where $\lambda_t$ is the energy clearing price (¥/MWh). $P_{\text{SG},t}^{\text{str}}$ is the output of strategic SGs. $O_{\text{SG}}^{\text{str}}$ is the marginal generation cost of strategic SGs (¥/MW). $O_{\text{BS}}^{\text{str}}$ is the levelized cost of BS owned by the strategic power company. $P_{\text{BS},t}^{\text{str}}$ is the charge/discharge power of BS correspondingly. Note that the cost of BS is expressed as second-order to avoid the minus sign when BS swift power flow [8]. $k_t$ is the strategic bidding decision. $k_{\max}$ is the maximum bidding decision. (2.1) indicates that revenues made by the power company are from the energy market clearing, the operating costs consider SGs, BSs, while WTs are considered as a free technique in this work. (2.2) confines the strategic bidding decisions.

## B. Lower Level Model

The SO in the LL problem takes the strategic energy offer $k_t O_{SG}^{str}$ and tries to clear the energy market at a minimum operation cost. The LL model can be expressed as follows:

$$\min \sum_t \left[ \sum_{SG^{str}} k_t O_{SG}^{str} P_{SG,t}^{str} + \sum_{BS^{str}} O_{BS}^{str} P_{BS,t}^{str\,2} + \sum_{SG^{n\text{-}str}} O_{SG}^{n\text{-}str} P_{SG,t}^{n\text{-}str} + \sum_{BS^{str}} O_{BS}^{n\text{-}str} P_{BS,t}^{n\text{-}str\,2} \right] \quad (3.1)$$

subject to:

$$\sum_{SG} P_{SG,t} + \sum_{BS} P_{BS,t} + \sum_{WT} P_{WT,t} = P_{D,t} : (\lambda_t) \quad (3.2)$$

$$0 \leq P_{SG,t} \leq P_{SG,max} \quad (3.3)$$

$$P_{SG,down} \leq P_{SG,t+1} - P_{SG,t} \leq P_{SG,up} \quad (3.4)$$

$$-P_{BS,max} \leq P_{BS,t} \leq P_{BS,max} \quad (3.5)$$

$$SoC_t = SoC_{t-1} - P_{BS,t}/E_{max} \quad (3.6)$$

$$SoC_{min} \leq SoC_t \leq SoC_{max} \quad (3.7)$$

$$SoC_0 = SoC_T \quad (3.8)$$

$$0 \leq P_{WT,t} \leq P_{WT,t} \quad (3.9)$$

where $P_{WT,t}$ is the output of WTs. $P_{D,t}$ is the system demand at period $t$. $P_{SG,up}$ and $P_{SG,down}$ are the ramp-up/ramp-down rates of SGs. $P_{BS,max}$ is the max charge/discharge power. $SoC_t$ is the State of Charge (SoC) of BSs at period $t$. $E_{max}$ is the capacity of BSs (MWh). $SoC_{max}$ and $SoC_{min}$ are the max and min SoC for BSs respectively. $SoC_0$ and $SoC_T$ are SoC of the start and end periods, respectively. $P_{WT,t}$ is the rated power of WTs at period $t$. (3.1) indicates the objective for minimizing the operating costs. (3.2) balances the supply and demand of the system. (3.3) and (3.4) are the output limits and ramp constraints of SGs respectively. (3.5)-(3.8) constrain the variables associated with BSs: (3.5) sets the bound of charging and discharging; (3.6) defines the SoC; (3.7)-(3.8) constrain the range of SoC, especially force the SoC at the end of the clearing cycle equal to the value of the first hour for securing that BS is able to effectively support the system in the next clearing cycle. (3.9) implies the output of WTs is within the rated capacity.

## C. Market Power Quantification Method

To quantify the market power of the strategic player, we refer to the define of Lerner index [9], then derive an averaged strategic bidding decisions in one-day operation. First, the expression of Lerner index is given as:

$$L = \frac{p - MC}{p} \quad (4.1)$$

where p represents the price of the good set by the firm and MC representing the true marginal cost of the firm considered.

In our work, (4.1) can be further expressed as:

$$L_t = \frac{k_t O_{SG}^{str} - O_{SG}^{str}}{k_t O_{SG}^{str}} = \frac{k_t - 1}{k_t} \quad (4.2)$$

TABLE I. PARAMETERS OF GENERATION SOURCES

| Parameters | CO-1 | CO-2 | CO-3 |
|---|---|---|---|
| $P_{SG,max}$ $P_{SG,up}$ $P_{SG,down}$ | 4, 2, 2 | 5, 2.5, 2.5 | 6, 3, 3 |
| $P_{SG,0}$ | 2 | 3 | 4 |
| $P_{BS,max}$ | 0.6 | 0.6 | 1.2 |
| $E_{max}$ | 1 | 1 | 2 |
| $SoC_0$ $SoC_{max}$ $SoC_{min}$ | 0.4, 0.9, 0.2 | 0.4, 0.9, 0.2 | 0.4, 0.9, 0.2 |
| $O_{SG}$ $O_{BS}$ | 900, 50 | 600, 50 | 500, 50 |

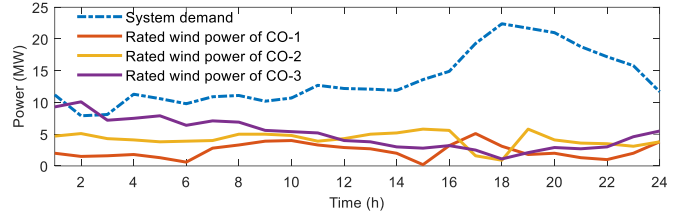

Figure 2. Profiles of load and rated wind power of companies.

TABLE II. EVALUATION OF ECONOMIC INDICATORS UNDER DIFFERENT MARKET COMPETITION CONDITIONS (UNIT IS K¥)

| Indicators | | PCM | ICM:CO-1 | ICM:CO-2 | ICM:CO-3 |
|---|---|---|---|---|---|
| Energy fee | | 37.87 | 69.14 | 79.61 | 75.66 |
| Profits | CO-1 | -5.83 | **2.15** | 2.17 | 0.34 |
| | CO-2 | -3.48 | 6.28 | **7.65** | 9.87 |
| | CO-3 | -2.45 | 11.09 | 17.75 | **14.79** |

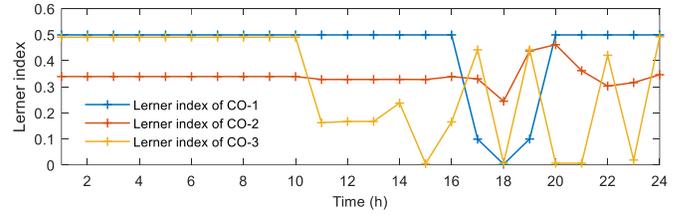

Figure 3. Lerner index evaluation of companies in one-day operation.

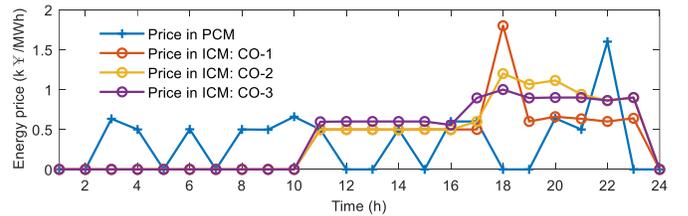

Figure 4. Energy clearing prices in different market conditions.

$$\bar{L} = \sum_t L_t / T \quad (4.3)$$

where (4.3) is the averaged version of Lerner index, only employing the strategic bidding decision. A large $\bar{L}$ means a strong market power for influencing prices.

## IV. CASE STUDY

The case studies in this paper use Julia as the programming language, and packages JuMP.jl, BilevelJuMP.jl to model the bi-level problem. Ipopt is chosen as the solver as it is capable of effectively tackling the proposed problem. The flexible resource parameters of three power companies (named as CO-1, CO-2, CO-3) are shown in Table I. The profiles of system demand and available wind power are shown in Fig. 2. The value of $k_{max}$ is set as 2.

### A. Analysis for Market Power of Strategic Companies

In this section, we focus on comparing the energy fee that consumers need to pay, profits of companies, energy clearing prices, and potential strategic bidding decisions in two markets, namely, Perfectly Competitive Market (PCM) and Imperfectly Competitive Market (ICM). The PCM case is a single-level problem that is optimized centrally; the ICM one is applied into three strategic scenarios depending on which company conducts strategic bidding.

*1) Analysis by Lerner Index:* The Lerner index for evaluating market power is illustrated in Fig. 3. Based on this one-day evaluation, we can find that the market power of CO-1 ($\bar{L} = 0.4447$) is stronger than CO-2 ($\bar{L} = 0.3402$) and CO-3 ($\bar{L} = 0.3108$). This is because CO-1 manages small-capacity, high-marginal-cost units that usually win bids when the balance between supply and demand is tight. The system relies heavily on its power generation (otherwise it cannot meet the load). Because they serve as the marginal units, they can raise the clearing price by overreporting marginal costs. For the CO-3, the market power is weak as it manages large-capacity units with low marginal costs. It is a base-load unit with a high probability of winning the bid and a small strategic space. If it raises bids, it is easy to lose the chance to win the bid, resulting in huge power generation losses. The lost opportunity cost is much higher than the gains obtained by trying to raise prices by falsely reporting. The marginal cost and capacity of SGs operated by CO-2 are between the above two, so the market power is also sandwiched between them.

At 18:00, all three companies tend to reduce market power, because peak hours are the time when electricity prices are the highest and most profitable. If the bid is overstated and no power is generated, an extremely valuable high-yield opportunity will be missed. Moreover, based on marginal energy pricing mechanism, even if the bid is low, a high marginal electricity price can be achieved. Subsequently, it is proved that no matter what the price submitted, at this hour, all SGs would reach the rated power to meet the load demand, the hard constraint, leading companies to make considerable profits.

*2) Analysis by Energy Prices:* The energy clearing prices under PCM (single-level problem) and ICM (bi-level problem) are shown in Fig. 4. When players are strategic, the market clearing price is 0 before 10:00, because the marginal bids of players during this period are higher than those under PCM, and the free wind power basically meets the load demand, so SGs are not dispatched. After the load increases (the available wind resources decrease meanwhile), the energy price rises accordingly, especially at 18:00, when the energy price reaches the highest level and all units reach the rated power, maintaining power balance while making profits. Since CO-1 has the highest marginal cost, followed by CO-2, when they each behave strategically, the cap of the energy price will also change accordingly to consider potential misreported prices. Therefore, at 18:00, the electricity prices in the three strategic scenarios are ranked as CO-1, CO-2, and CO-3. The PCM determines almost the opposite price, especially the electricity price at 18:00 is 0, indicating the differences in market clearing under different market conditions.

*3) Analysis by Profits and Payment:* The electricity purchase costs for consumers and profits of each power company under different scenarios are given in Table II. It can be analyzed that compared with the situation where each company submits the real marginal price to clear the market, i.e., PCM, when there are strategic players in the market, i.e., ICM, the electricity purchase cost of users will increase significantly, from 37.87 k¥ to 69.14 k¥, almost doubling. Correspondingly, the profits of each company also increase significantly, from negative to positive. This proves that *the strategic behavior of selfish power companies will make a lot of profits with causing consumers to bear more electricity bills*.

In order to deal with the huge bills caused by the strategic behavior of power companies, potential measures to mitigate market power can be considered, including limits of the bidding decision, i.e., $k_{max}$ or uniformly setting a clearing price cap. Both of them are conducted by market regulators, but the difficulty lies in how to determine the real marginal prices of units to reasonably limit the strategic space.

### B. Analysis for Reserve Capacity under Scenarios

The capacities of reserve-I and reserve-II in one-day operation under different market conditions (PCM, ICM:CO-1, CO-2, CO-3) are illustrated in Fig. 5 and 6 respectively. The corresponding averaged capacity is indicated in Table. III. Since SGs are not dispatched before 10:00, both reserve capacities are zero. We can clearly see that, *without loss of generality, an imperfectly competitive market can enable the system to have more spinning reserve capacity*. For both types of reserve, generally, at 19:00-21:00 and 23:00, the reserve under ICM is greater than those under PCM, especially at 20:00-21:00, the system reserve under PCM is zero, which may cause unstable system operation. In addition to the above time periods, when the CO-3 strategically bids, the reserve capacity at 15:00-16:00 also increases significantly.

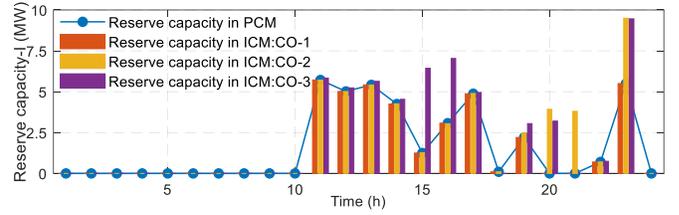

Figure 5. Reserve capacity type-I in various scenarios.

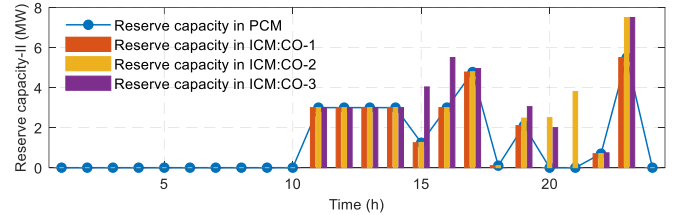

Figure 6. Reserve capacity type-II in various scenarios.

TABLE III. AVERAGED RESERVE CAPACITY UNDER VARIOUS CONDITIONS (UNIT IS MW)

| Reserve type | PCM | ICM:CO-1 | ICM:CO-2 | ICM:CO-3 |
|---|---|---|---|---|
| Type-I | 1.59 | 1.59 | 2.09 | 2.34 |
| Type-II | 1.23 | 1.23 | 1.59 | 1.66 |

The reason for the above is that when there is market power, the SO may be 'forced' to dispatch high-priced SGs, even if they have a high cost, they will enter the clearing with operating at a minimum output point, and as a result, more potential reserve capacity will be retained. Under PCM conditions, all SGs bid at marginal cost, and SO chooses the lowest cost combination to clear to meet the load. If wind power meets the load, some SGs may be shut down during dispatch. Although this leads to lower electricity prices, there are fewer SGs online and less remaining reserve capacity. On the contrary, when market power exists, low-cost SGs falsely bid high prices or fail to clear. SO has to select some SGs to meet the load. The output of the SGs themselves may not be the maximum, but after being turned on, it leaves the system with considerable reserve capacities. In addition, as the energy clearing price is raised by strategic SGs, other non-strategic units may also be motivated to start up to earn profits and provide reserve. The final result is: although the energy payment is higher, there are more online SGs, and the reserve capacity is enhanced.

## V. CONCLUSION

Stakeholders in the electricity market generally do not make real bids to make profits, and selfishly affect the market energy clearing. Meanwhile, the increasing penetration of RESs pose challenges to the stability of the grid, and a certain amount of spinning reserve capacity needs to be provided to ensure a safe system operation. In the above context, this paper uses a bi-level model to simulate the imperfectly competitive market and explore the impact of selfish behaviors on reserve capacity. To derive the KKT conditions, we propose an assumption for offline calculation of reserve capacity, which makes the model continuous and solvable.

Case studies show that in the imperfectly competitive market, more units are dispatched online, and energy prices are likely to be cleared as high, which will increase electricity bills of the consumers. But the reserve capacity of the entire system can potentially be increased. Future work should focus on measuring the balance between consumer bills and system reserve capacity to build an economical and safe system.